\UseRawInputEncoding 
\documentclass{ws-ijbc}

\usepackage[normalem]{ulem}

\usepackage{enumitem}
\usepackage{multicol}
\usepackage{ws-rotating}     
\usepackage{graphicx}
\usepackage{epstopdf}
\usepackage{tabularx}
\usepackage[titlenumbered,ruled]{algorithm2e}
\usepackage{listings}
\usepackage{lmodern}
\usepackage[usenames,dvipsnames]{xcolor}

\usepackage{algorithmicx}
\usepackage{algpseudocode}
\usepackage{xcolor}

\usepackage[toc,page]{appendix}

\algrenewcommand\alglinenumber[1]{\tiny #1:}
\usepackage{mathtools}
\usepackage{amsmath}%
\usepackage{amsfonts}%
\usepackage{amssymb}%
\usepackage{graphicx}
\usepackage{color}
\usepackage{cases}
\def\HiLi{\leavevmode\rlap{\hbox to \hsize{\color{yellow!50}\leaders\hrule height .8\baselineskip depth .5ex\hfill}}}

\newlist{myenumi}{description}{10}
\setlist[myenumi]{labelindent=\parindent, leftmargin=*, label=(\roman*), align=left}
\setlist[myenumi]{leftmargin=0pt}

\renewcommand{\labelitemi}{$-$}

\definecolor{darkgreen}{rgb}{0.1, 0.5, 0.1}
\begin{document}

\markboth{M.-F. Danca}{Matlab code for Lyapunov exponents of fractional-order systems, Part II}

\title{Matlab code for Lyapunov exponents of fractional-order systems, Part II: The non-commensurate case}

\author{MARIUS-F. DANCA}

\address{Romanian Institute od Science and Technology,\\
400487 Cluj-Napoca, Romania,\\
danca@rist.ro}

\maketitle

\begin{history}
\received{(to be inserted by publisher)}
\end{history}

\begin{abstract}

In this paper, the Benettin-Wolf algorithm for determining all Lyapunov exponents of non-commensurate fractional-order systems modeled by Caputo\textsc{\char13}s derivative and the corresponding Matlab code are presented. The paper continues the work started in \cite{dadus}, where the Matlab code of commensurate fractional-order systems is given. To integrate the extended systems, the Adams-Bashforth-Moulton scheme for fractional differential equations is utilized. Like the Matlab program for commensurate-order systems, the program presented in this paper prints and plots all Lyapunov exponents as function of time. The program can be simply adapted to plot the evolution of the Lyapunov exponents as a function of orders, or a function of a bifurcation parameter. A special attention is paid to the periodicity of fractional-order systems and its influences. The case of non-commensurate the Lorenz system is demonstrated.
\end{abstract}



\section{Introduction}
\begin{multicols}{2}

Despite the fact that there are opinions and results questioning the utility of Lyapunov Exponents (LEs)(see e.g. \cite{ChaosBook}, where the evaluation of the LEs is not recommended: ``Compute stability exponents and the associated covariant vectors instead. Cost less and gets you more insight. ... we are doubtful of their utility as means of predicting any observables of physical significance''), determining numerically LEs remains the subject of many works growing into a real software industry for modern nonlinear physics
(see, e.g.\cite{HeggerKS-1999,BarreiraP-2001,Skokos-2010,CzornikNN-2013,PikovskyP-2016,VallejoS-2017} and others).

In numerical calculations of the LEs, the asymptotic time averaging is usually accomplished by using a sufficiently long time to allow the average of the exponents to converge within a set tolerance. Although less frequently used, probability densities (distributions) of the exponents, averaged over a much shorter time, also contain valuable dynamic information. Such distributions are made up of the so-called finite-time or local LEs \cite{ot,pras} (see also \cite{bot}).

The global Lyapunov exponent can measure the time-averaged divergence of nearby trajectories on a strange attractor, revealing an exponential divergence (in the case of positive exponents) or convergence (in the case of negative exponents) \cite{ruel,lyappp}. The largest Lyapunov exponent gives a rough estimate of the predictability limit \cite{huai}. On the other side, numerically there often exists a strong interest in the local  predictability. In ergodic systems, most trajectories will asymptotically yield to the
same LE. If trajectories are considered for short times only, then the mean separation rate will depend on the trajectories and also the length of the time interval \cite{gras,br}.
Since finite time LEs can be obtained as integrals of the local separation rate along the trajectory, they are also called local LEs \cite{br}.
Approaches for the LEs computation and their differences are discussed, e.g., in \cite{KuznetsovAL-2016,kuz1}.

Nowadays, there are two widely used definitions of the LEs: via the exponential growth rates of norms of the fundamental matrix columns  \cite{Lyapunov-1892}
and via the exponential growth rates of the sigular values of the fundamental matrix
\cite{Oseledec-1968}.

Applying the statistical physics approach and assuming the ergodicity
(see, e.g. \cite{Oseledec-1968}),
LEs of a given dynamical system are often estimated
by local LEs along a ``typical'' trajectory.
However, in numerical experiments,
the rigorous use of the ergodic theory is a challenging task
(see, e.g. \cite[p.118]{ChaosBook}).

Since in numerical experiments only the finite time LEs can be computed, they can differ significantly from the limit values, such as in the case where the considered trajectory belongs to a transient chaotic set. Therefore, it is remarkable that these characteristic quantities are reliable to only a few decimals.

Although the subject of fractional derivative is three centuries old as the conventional integer-order calculus, is the use of this kind of operators mainly started recently (see e.g. \cite{cap}). Nowadays, differential or difference equations of FO represent useful models in mechatronics, viscoelasticity, seismology, electrical circuits, aerodynamics, biophysics, biology, blood flow phenomena, chemistry, control theory, etc. (see, e.g., \cite{mac} or references in \cite{tava2}).

Analytical solutions for fractional differentiation problems are very limited. Due to this reason, recently many numerical approximations of solutions to fractional equations were proposed (see e.g. \cite{kai}).

On the other side, a recent result opened a lot of questions on the applicability of fractional differential operators: nonexistence of exact non-constant periodic solutions of autonomous nonlinear dynamical systems of FO, proved by Tavazoei and Haeri in 2009 \cite{tava}. Once this result appeared, a lot of related works have been published (see for example \cite{area,yaz,kan}).
The impact of this result in the applications of Fractional Order (FO) systems seems to be more important than his influence in the theory of FO systems. Therefore, while the number of theoretic results on this subject continues to born, non-periodicity affect the rightness of many applicative works on FO systems, appeared after this result.

Therefore, until some analytical modality to overcome this problem appears, in this paper a possible way to avoid this problem of non-periodicity is described.

\section{On the periodicity of numerical solutions of fractional-order systems}\label{sec2}

As mentioned in Introduction, one of the most important property of nonlinear systems, periodicity, is impossible in FO systems, discrete or continuous.

Consider the Initial Value Problem (IVP) of (non-commensurate or commensurate) of FO, with Caputo's derivative:
\begin{equation}\label{unu}
\begin{array}{l}
D_*^qx(t)=f(x(t)),\\
x(0)=x_0,
\end{array}%
\end{equation}

\noindent for $t\in[0,T]$, $q\in(0,1)$, $f:\mathbb{R}^n\rightarrow \mathbb{R}^n$ and $D_*^q$, Caputo's differential operator of order $q$ with starting point $0$:

    \begin{equation*}
    D_*^qx(t)=\frac{1}{\Gamma(1-q)}\int_0^t(t-\tau)^{-q}x'(\tau)d\tau,
    \end{equation*}
    with $\Gamma$ being the Euler function.

Properties of Caputo's differential operator, $D_*^q$, can be found in e.g. \cite{pod2,gore}.

\begin{lemma}\cite{tava}\label{t1} Systems modeled by the IVP \eqref{unu} can not have any non-constant periodic solution.
\end{lemma}

This result is based on the result which states that for $q\in(0,1)$, Caputo's derivative of a non-constant $T$-periodic function cannot be $T$-periodic (as for Lemma \ref{t1}, the result has been proved also for Grunwald-Letnikov and Riemann-Liouville derivatives) and is due to the fact that the entire past history of the system has to be taken into account.

An extended result of Lemma \ref{t1} is the following theorem

\begin{theorem}\cite{yaz}\label{t2} Systems modeled by the IVP \eqref{unu} does not have any periodic solution unless the lower terminal of the derivative is $-\infty$.
\end{theorem}

\begin{figure}
\begin{center}
\includegraphics[scale=0.75]{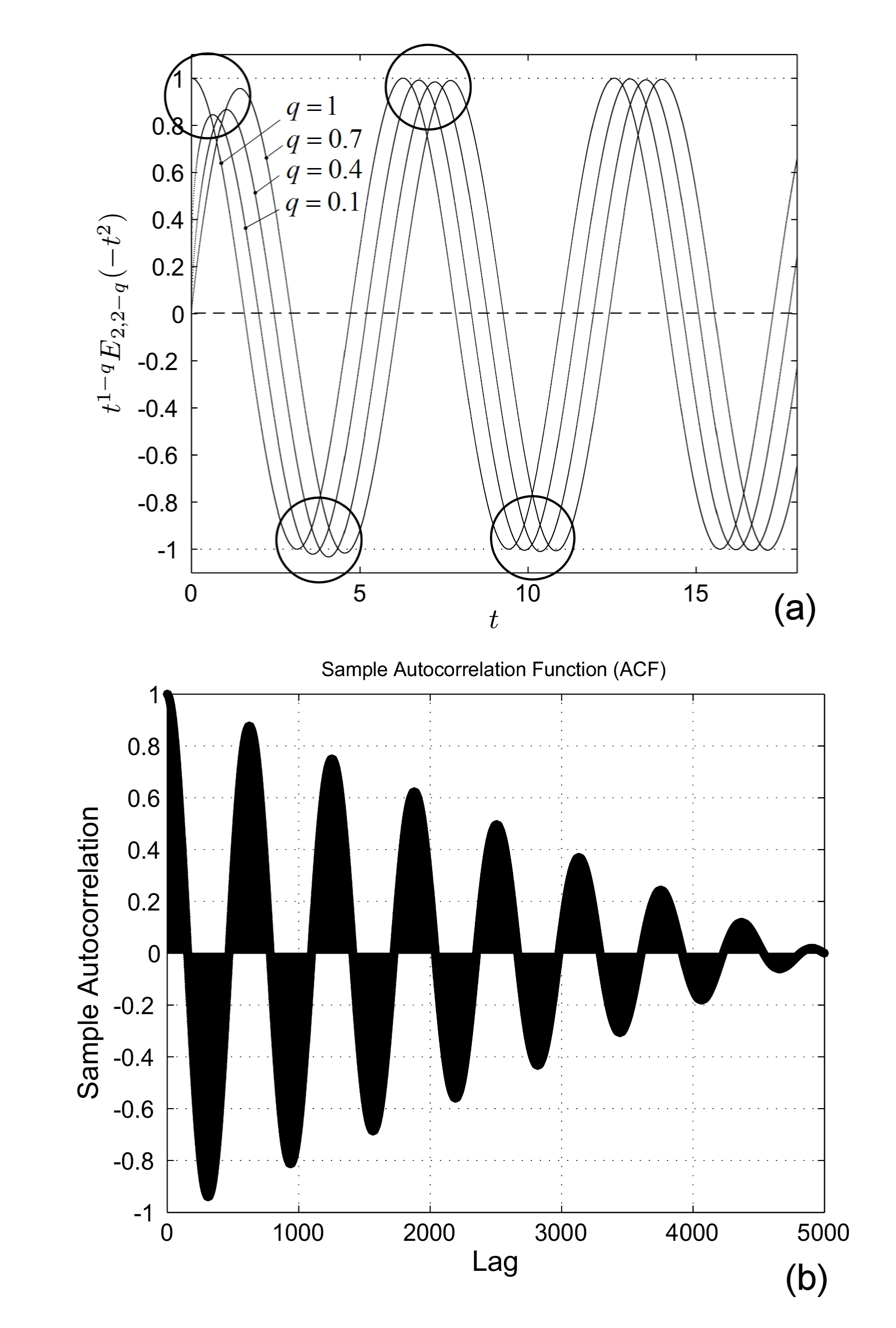}
\caption{(a) Graphs of the derivative $D_*^q \sin(t)=t^{1-q}E_{2,2-q}(-t^q)$ for $q\in\{0.1,0.4,0.7,1\}$. Circled regions reveal the differences between the derivative of IO of $\sin$ ($q=1$), and the non periodic derivatives for $q\in\{0.1,0.4,0.7\}$; (b) Autocorrelation of the derivative $D_*^q$ for $q=0.4$.}
\label{fig1}
\end{center}
\end{figure}

Therefore, just the fractional systems with the lower terminal of $-\infty$ could have periodic solutions.
To understand better Lemma \ref{t1}, revisit the example of Caputo's derivative of the periodic sine function, whose Caputo's derivative for $q\in(0,1)$ is not periodic (In \cite{tava2}, the example is introduced via Riemann-Liuville derivative).
\[
D_*^q \sin(t)=t^{1-q}E_{2,2-q}(-t^q),
\]
where $E_{a,b}(z)$ is the two-parameter function of Mittag-Leffler \cite{poduss,pod2}
\[
E_{a,b}(z)=\sum_{k=0}^\infty \frac{z^k}{\Gamma(ak+b)}.
\]
Although sine is a periodic function, its fractional derivative, $D_*^q \sin(t)$, is not, compared with the integer-order derivative, when $\frac{d}{dt}\sin(t)=\cos(t)$, which is periodic. In Fig. \ref{fig1} (a), the function $t^{1-q}E_{2,2-q}(-t^2)$ for 4 values of $q$ is presented. The circled regions show that except the case $q=1$ when the function $t^{1-q}E_{2,2-q}(-t^2)$ becomes the periodic function $\cos(t)$ (\cite{pod2} Theorem 1.7), for $q\in\{0.1,0.4,0.7\}$ the underlying graphs are not periodic. However, because usually first transients have to be removed, the best way to show numerically the aperiodicity of $D_*^q\sin(t)$ for larger values of $t$ (as necessary for, e.g., stable cycles), is to calculate the autocorrelation which, for non-periodic time series (signals), decreases to zero for relatively large values of time (see Fig. \ref{fig1} (b) for $q=0.4$).

Another interesting result refers to the special case of asymptotic stability, namely the finite-time stability of FO systems. The equilibrium $x = x^*$ of system \eqref{unu} is said to be (locally) finite-time stable if it is stable and, for the trajectory $x(t)$ starting from $x_0$ located in a neighborhood of $x^*$, there exists a time instant $T > 0$, such that $x(t) = x^*$ for all $t \geq T$.

\begin{theorem}\cite{lam} If $x=x^*$ is an equilibrium of the IVP \eqref{unu}, then $x=x^*$ cannot be finite-time stable.
\end{theorem}

Some of the practical implications of above results are clearly underlined in \cite{tava}: ``Can
the controller tuning methods or identification methods, which
use limit cycle information, be extended for use in the control or
identification of FO systems? As another example, let
us consider the issue of modeling. We know that
differential equations OF fo (FDEs) have been effectively used in the modeling
of real-world systems. According to the result presented in this
note, which model structure should be preferred for modeling
an oscillatory system, an IO or a FO? Are
Caputo based FDEs, which do
not have periodic solutions, good candidates to model oscillatory
systems? Let us give another example. Stabilization of unstable
periodic orbits as a control objective is a way to suppress the
chaotic oscillations in IO chaotic systems. To achieve this goal, many control techniques
have been proposed, such as the OGY method or the Pyragas method, which can
reduce the chaotic oscillations to regular oscillations. Now, this
question may be struck. How to reduce the chaotic oscillations
to regular oscillations in chaotic FO systems, while
we know that FO chaotic systems do not have
periodic orbits? Another related question is about the dynamical
analysis of FO chaotic systems. It has been found that
unstable periodic orbits play an important role in understanding
the complex behavior of IO chaotic systems. However,
how do we understand and analyze the behavior of fractional
order chaotic systems, while there exists no periodic orbit in these
systems?''

Also, other problems regard: a) bifurcation diagrams where, beside chaotic windows, periodic windows can appear; b) many chaotic attractors in FO systems, embedding the set of unstable periodic trajectories which form a dense set; c) Hopf theorem which implies limit cycles; d) synchronization of chaotic FO systems, with chaos containing unstable periodic trajectories; e) finite-time synchronization of fractional-order chaotic systems via terminal sliding mode control based on periodic motions.

Considering the above results, it is clear that FO systems can have only asymptotically non-constant periodic solution and, moreover, to our knowledge nothing is known about the stability of this kind of solutions.

One over the other, many phenomena and real systems are not strictly periodic and, before a rigorous theoretical answer to the above open problems will appear, in order to overcome this obstacle, consider the following definition.

\begin{definition}\label{defa}In the $n$-dimensional phase space $\mathbb{R}^n$, with $n\geq2$, a numerically periodic trajectory (NPT) refers to as a closed trajectory in the numerically sense that, after transients are removed, the closing error $\epsilon$ is within a given bound of $1E - m$, with $m$ being a sufficiently large positive integer (see the sketch in Fig. 2).
\end{definition}

For convenience,  we  assume  from  now on that even no results on the stability of asymptotically periodic FO orbits are known, NPTs can be considered as good candidates for numerical approaches.

\section{Matlab code for Lyapunov exponents}
The following result provide the existence of the LEs of the FO system modeled by the IVP \eqref{unu} \cite{exist}

\begin{theorem}
System \eqref{unu} has the following variational equations, which define the LEs:
\begin{equation}\label{doi}
\begin{array}{l}
D_*^q \Phi(t)=D_xf(x)\Phi(t),\\
\Phi(0)=I,
\end{array}
\end{equation}
\noindent where $\Phi$ is the matrix solution of system \eqref{unu}, $D_x$ is the Jacobian of $f$ and $I$ is the identity matrix.
\end{theorem}

The utilized algorithm for all LEs of FO systems, has been proposed in the seminal works of Benettin et al. \cite{bene2} (see also \cite{shima}), one of the first work to propose a Gram-Schmidt orthogonalization procedure to compute LEs for continuous systems of IO, as described in \cite{lili}), and by Wolf et al. \cite{lyappp} (see also \cite{eci}).
While in the previous paper, the Matlab code to determine the LEs is based on the PECE numerical ABM, for integration of the IVP \eqref{unu} of commensurate order \cite{kai} now, the numerical integrator is the same method, but for non-commensurate order.

The algorithm to find all LEs of Integer Order (IO) from a time series, is described as a Fortran code and later as a Matlab code (see \cite{lyappp} and \cite{alan}, respectively).

The program presented in this paper uses the skeleton of the Matlab code presented in \cite{dadus}, based on the program lyapunov.m (see \cite{govo}).

While in \cite{dadus}, the required numerical integration of the commensurate order system uses a numerical method for FDEs of commensurate order, in this paper the algorithm remains basically similar. But, instead of a numerical method for FDEs of commensurate-order, a numerical method for FDEs of non-commensurate order is utilized. Note that for both commensurate and non-commensurate order cases, any other fast method to integrate the IVP \eqref{unu} can be used.

Because numerical integration of the IVP \eqref{unu} is time consuming, it is preferable to employ improved numerical code, like the FDE12 used in \cite{dadus}. In this paper, the program \texttt{fde\_pi12\_pc.m} \cite{robi} (see also \cite{robi2}) is used.

The code for FO LEs of non-commensurate order, called \texttt{FO\_nc\_Lyapunov.m} (see Appendix \ref{a} where, for speed, starred lines can be put as comment), uses a similar code for commensurate order, the file containing the extended system, \texttt{ext\_fcn.m}, and a solver for non-commensurate FDEs (here, \texttt{fde\_pi12\_pc.m}). All files must be in the same folder.

The running command of \texttt{FO\_nc\_Lyapunov.m} is

\begin{lstlisting}
 [t,LE]=FO_NC_Lyapunov(ne,ext_fcn,t_start,h_norm,t_end,x_start,h,q,out);
\end{lstlisting}
where
\begin{itemize}[leftmargin=.2in]
\item \texttt{ne} represents the equations (and state variables) number;
\item \texttt{ext}$\_$\texttt{fcn.m} the function containing the extended system;
\item \texttt{t}$\_$\texttt{start} and \texttt{t}$\_$\texttt{end} the time span;
\item \texttt{h}$\_$\texttt{norm} the normalization step in the Gram-Schmidt algorithm;
\item \texttt{x}$\_$\texttt{start} the initial condition;
\item \texttt{h} the integration step size;
\item \texttt{q}=\texttt{[q}$_1$,\texttt{q}$_2$,...,\texttt{q}$_\texttt{ne}$], and
\item \texttt{out} indicates the number of \texttt{h\_norm} steps when intermediate values of time and LEs are printed (for \texttt{out=0}, no intermediate results will be printed out).
\end{itemize}

Consider the Lorenz system of noN-commensurate order:

\[\label{lorus}
\begin{array}{l}
D_*^{q_1} x_1=a(x_2-x_1),\\
D_*^{q_2} x_2=x_1(b-x_3)-x_2,\\
D_*^{q_3} x_3=x_1x_2-cx_3,
\end{array}
\]
with $b$ variable, $a=10$, $c=\frac{8}{3}$, the FO $q=[q_1,q_2,q_3]=[0.995,0.992,0.996]$ and the extended function, \texttt{Lorenz\_ext.m}, presented in Appendix \ref{b}.

For $b=23$, one of the two equilibria $(\pm\sqrt{c(b-1)}, \pm\sqrt{c(b-1)},b-1)$, $(7.659,7.659,22.000)$, attracts the trajectory starting from the initial condition \texttt{x\_start=[1,1,1]} (Fig. \ref{fig3} (a)).\footnote{ The use of the code can be simplified so that the parameter (here $b$) be set from the command line of \texttt{FO\_nc\_Lyapunov} (see \cite{dadus}).}

The LEs are obtained with the command
\begin{lstlisting}
[t,LE]=FO_NC_Lyapunov(3,@Lorenz_ext,0,0.1,1000,...
[1,1,1]',0.01,[0.995,0.992,0.996],1000)
\end{lstlisting}

and are presented in Table \ref{tab1} with the time-evolution in Fig. \ref{fig3} (b).

\lstset{%
    caption=Caption Text,
    frame=tb
  }
\begin{lstlisting}[label=tab1]
    100.00    -0.0637    -0.0962   -13.6300
    200.00    -0.0679    -0.0861   -13.6367
    300.00    -0.0705    -0.0816   -13.6389
    400.00    -0.0711    -0.0800   -13.6400
    500.00    -0.0718    -0.0787   -13.6407
    600.00    -0.0722    -0.0779   -13.6412
    700.00    -0.0724    -0.0774   -13.6415
    800.00    -0.0727    -0.0769   -13.6417
    900.00    -0.0727    -0.0767   -13.6419
   <@\textcolor{blue}{1000.00}@>   <@\textcolor{blue}{-0.0730}@>  <@ \textcolor{blue}{-0.0763} @> <@\textcolor{blue}{-13.6421}  @>
\end{lstlisting}
Table 1. LEs of the Lorenz system \eqref{lorus} for $b=23$ (last blue line).
\vspace{3mm}
\lstset{%
    frame=single
  }

For $b=28$, the system evolves chaoticALLY (Fig. \ref{fig4} (a)) and the LEs are given in Table \ref{tab2} with the time evolution in Fig. \ref{fig4} (b).

\lstset{%
    caption=Caption Text,
    frame=tb
  }
\begin{lstlisting}[label=tab2]
    100.00     0.7637     0.0026   -14.5118
    200.00     0.8306     0.0022   -14.5751
    300.00     0.8650     0.0033   -14.6086
    400.00     0.8854    -0.0022   -14.6226
    500.00     0.8801     0.0042   -14.6238
    600.00     0.8811     0.0024   -14.6233
    700.00     0.8861     0.0007   -14.6264
    800.00     0.8918     0.0010   -14.6320
    900.00     0.8961    -0.0008   -14.6344
   <@\textcolor{blue}{1000.00}@>   <@  \textcolor{blue}{0.8956}@>   <@    \textcolor{blue}{0.0006} @> <@\textcolor{blue}{-14.6352}  @>
\end{lstlisting}
Table 2. LEs of the Lorenz system \eqref{lorus} for $b=28$ (last blue line).
\vspace{3mm}
\lstset{%
    frame=single
  }

Considering the results and comments in Section \ref{sec2}, the above results should be considered in the spirit of Definition \ref{defa}. Thus, for $b=23$, the negativeness of LEs would indicate the existence of a NPT.

Note that because of the numerically approximation of the NPTs, beside the errors typically due to numerically algorithms for LEs, it is possible to have supplementary errors.

\begin{remark}
With minor modifications, the code can be used to plot the evolution of LEs vs some system parameter or vs \texttt{q} (see \texttt{FO\_Lyapunov\_p.m}, \texttt{run\_FO\_Lyapunov\_p.m}, and \texttt{FO\_Lyapunov\_q.m}, \texttt{run\_FO\_Lyapunov\_q.m}, respectively, in \cite{dadus}).
\end{remark}

As mentioned in \cite{dadus}, the main steps to determine numerically the LEs are: numerical integration of the FO system \eqref{unu} together with the variational system \eqref{doi} (i.e. the extended system), and the correlation between \texttt{h\_norm} of the Gram-Schmidt procedure and \texttt{h} of the numerical integration. Thus, beside initial conditions, one of the most important parameters is \texttt{h\_norm} and also its relation with \texttt{h} (Fig. \ref{fig5}). Thus, while the role of initial conditions especially in the case of chaotic behavior, when LEs present a strong sensitivity dependence on initial conditions, is well known and studied, the influence of the size of \texttt{h$\_$norm}, not only for the FO case but also in the IO case, is not well established from the numerical point of view. Obviously, correct results can be obtained only if \texttt{h\_norm} is multiple of \texttt{h}, but several tests with different (\texttt{h}, \texttt{h\_norm} ) have to be tried for a specific system until the obtained LEs present an invariance-like result with \texttt{h} and \texttt{h\_norm}.

\section*{Conclusion and discussion}
In this paper, the Matlab code for determining numerically the LEs of a class of dynamical systems of non-commensurate order is presented. The paper continues the previous paper \cite{dadus}, where the case of commensurate order is analyzed. The code is designed by modifying the Matlab code presented in \cite{dadus}. A fast Matlab code to integrate non-commensurate FO systems is utilized. However, any other fast routines can be tried to integrate the extended system. A special attention should be paid to the relation between $h_norm$ of the Gram-Schmidt orthogonalization and the step $h$ of the integration routine. Without some general rule found, we suggest the relation \texttt{h\_norm=k$\times$ h} with \texttt{k} being some positive integer, e.g. $k=10$, but with values of these two parameters adequately chosen for every system.

Regarding the significance of LEs, note that the simple interpretation of positiveness as a sufficiency criteria for chaos is not always true. For example, in systems of IO with Perron effects, positive LEs may not imply instability and chaos \cite{dadus}. We strongly believe that this effect might arise from systems of FO too.

Since the routine \texttt{fde\_pi12\_pc.m} can be used for order $q=1$ (see \cite{robi2}), the code proposed in this paper can be used for systems modeled by equations of FO and IO. Actually, in order to reduce the mentioned inherent errors for a given system to which one know the LEs for the IO case, one can run the code \texttt{FO\_nc\_Lyapunov.m} with $q=1$ and adjust the above parameters \texttt{h} and \texttt{h\_norm} until the obtained LEs are approximatively identical with the IO case. Also, an interesting study would be to compare the results obtained with algorithms from time series with the results obtained using the equations of the model such as the presented algorithm.

As the author of the utilized routine \texttt{lyapunov.m} doesn't give any warranty regarding the accuracy of the results (a normal assumption for all algorithms for LEs), we advise readers to use the two proposed codes, the one for the commensurate case \cite{dadus} and the present code, with a justified precaution. In the author's opinion, it is almost impossible to write a 100\text \% efficient program for finding LEs of IO or FO systems by using whatever algorithm. Therefore, finding numerically LEs remains a more spectacular than precise tool in the studies of nonlinear dynamics.

Finally, it is noted beside the caution required by the relative accuracy of the numerically obtained values of LEs, typically for all numerical methods for LEs, a special attention should be given to the periodicity problem of FO systems presented in Section \ref{sec2}.

The code can be downloaded from MATLAB Central File Exchange:

https://www.mathworks.com/matlabcentral/\\fileexchange/92753-code-for-noncommensurate-fractional-order-lyapunov-exponents


\newpage{\pagestyle{empty}}

\section*{Appendices}

\addcontentsline{toc}{section}{Appendices}
\renewcommand{\thesubsection}{\Alph{subsection}}
\subsection{\textup {Matlab code for LEs of non-commensurate FO}}\label{a}

\begin{lstlisting}%[multicols=2]%[    basicstyle=\footnotesize, %or \small or \footnotesize etc.]

function [t,LE]=FO_NC_Lyapunov(ne,ext_fcn,t_start,h_norm,t_end,x_start,h,q,out);
%
% Program to compute the spectrum of Lyapunov
% exponents as function of time of systems
% of non-commensurate fractional order defined
% with Caputo's derivative
% ================================================
% author Marius-F. Danca
% web:http://danca.rist.ro/
% email: danca@rist.ro
% =================================================
% The program uses a fast variant of the Adams-
% Bashforth-Moulton for fractional-
% order differential equations: fde_pi12_pc.m, by
% Roberto Garrappa: shorturl.at/izJW3,
% see also
% R.Garrappa,  Numerical Solution of
% Fractional Differential Equations: A Survey and
% a Software Tutorial, Mathematics 2018, 6(2), 16.
%
% files required: fde_pi12_pc, FO_NC_Lyapunov.m
% and the function containing the extended system,
% ext_fnc.
% Files FO_NC_Lyapunov.m, fde_pi12_pc.m and
% ext_fnc must be in the same folder.
%
% FO_NC_Lyapunov.m is developed, by modifying
% the program for the commensurate order
% FO_Lyapunov.m: shorturl.at/suzF7 (see also
% Lyapunov.m by V.N. Govorukhin)
%
% Input:
% ne - system dimension;
% ext_fcn -function containing the extended system;
% t_start, t_end - time span for fde_pi12_pc.m;
% h_norm - step for Gram-Schmidt renormalization;
% x_start - initial condition;
% h - integration step;
% q=[q_1;q_2;...;q_ne] - the fractional order;
% out - priniting step of LEs values;
% out==0 - no print.
%
% Output:
% t - time values;
% LE Lyapunov exponents to each time value printed
% every 'out'*h_norm steps.
%
% How to use it:
%
% [t,LE]=FO_NC_Lyapunov(ne,ext_fcn,t_start,h_norm,
% t_end,x_start,h,q,out);
%
% Example:
%
% [t,LE]=FO_NC_Lyapunov(3,@Lorenz_ext.m,
% 0,0.1,1000,[1,1,1]',0.01,[.995,.992,.996],1000)
%
% For speed, starred lines can be commented.
%
%==================================================
% Cite the code as:
%
% Marius-F. Danca, Matlab code for Lyapunov 
% exponents of fractional-order systems, 
% Part II: The non-commensurate case, IJBC, 
% accepted on Jun 12, 2021.
%
% The algorithm for COMMENSURATE order is
% explained in:
%
% [1] Marius-F. Danca, Nikolay Kuznetsov, Matlab
% code for Lyapunov exponents of fractional order
% systems, IJBC, 28(05), 1850067 (2018)
%
% =================================================
figure;
hold on;
% Set orders q for extended system (ne system
% equations + ne*ne variational equations)
 q=repmat(q',ne+1,1);
% Memory allocation
x=zeros(ne*(ne+1),1);
x0=x;
c=zeros(ne,1);
gsc=c; zn=c;
n_it = round((t_end-t_start)/h_norm);
% Initial values for extended system
x(1:ne)=x_start;
i=1;
while i<=ne
    x((ne+1)*i)=1.0;
    i=i+1;
end
t=t_start;
LE=zeros(ne,1);
% Main loop
it=1;
while it<=n_it
% Solution of extended ODE system on [t,t+h_nrom]
    [T,Y] = fde_pi12_pc(q,ext_fcn,t,t+h_norm,x,h);
    t=t+h_norm;
    Y=transpose(Y);
    x=Y(size(Y,1),:); %solution at t+h_norm
    i=1;
    while i<=ne
        j=1;
        while j<=ne;
            x0(ne*i+j)=x(ne*j+i);
            j=j+1;
        end;
        i=i+1;
    end;
%   construct new orthonormal basis by gram-schmidt
    zn(1)=0.0;
    j=1;
    while j<=ne
        zn(1)=zn(1)+x0(ne*j+1)^2;
        j=j+1;
    end;
    zn(1)=sqrt(zn(1));
    j=1;
    while j<=ne
        x0(ne*j+1)=x0(ne*j+1)/zn(1);
        j=j+1;
    end
    j=2;
    while j<=ne
        k=1;
        while k<=j-1
            gsc(k)=0.0;
            l=1;
            while l<=ne;
                gsc(k)=gsc(k)+x0(ne*l+j)*x0(ne*l+k);
                l=l+1;
            end
            k=k+1;
        end
        k=1;
        while k<=ne
            l=1;
            while l<=j-1
                x0(ne*k+j)=x0(ne*k+j)-gsc(l)*x0(ne*k+l);
                l=l+1;
            end
            k=k+1;
        end;
        zn(j)=0.0;
        k=1;
        while k<=ne
            zn(j)=zn(j)+x0(ne*k+j)^2;
            k=k+1;
        end
        zn(j)=sqrt(zn(j));
        k=1;
        while k<=ne
            x0(ne*k+j)=x0(ne*k+j)/zn(j);
            k=k+1;
        end
            j=j+1;
    end
%   update running vector magnitudes
    k=1;
    while k<=ne;
        c(k)=c(k)+log(zn(k));
        k=k+1;
    end;
%   normalize exponent
    k=1;
    while k<=ne
        LE(k)=c(k)/(t-t_start);
        k=k+1;
    end
    i=1;
    while i<=ne
        j=1;
        while j<=ne;
            x(ne*j+i)=x0(ne*i+j);
            j=j+1;
        end
        i=i+1;
    end;
        x=transpose(x);
        it=it+1;
%       print and plot results
        if (mod(it,out)==0)%                   <@\textcolor{red}{(*)}@>
            fprintf('%10.2f %10.4f %10.4f %10.4f\n',[t,LE']);                        <@\textcolor{red}{(*)}@>
        end;%                                  <@\textcolor{red}{(*)}@>
        plot(t,LE)%                            <@\textcolor{red}{(*)}@>
end
xlabel('t','fontsize',16)%                     <@\textcolor{red}{(*)}@>
ylabel('LEs','fontsize',14)%                   <@\textcolor{red}{(*)}@>
set(gca,'fontsize',14)%                        <@\textcolor{red}{(*)}@>
box on;%                                       <@\textcolor{red}{(*)}@>
line([0,t],[0,0],'color','k')%                 <@\textcolor{red}{(*)}@>
axis tight%                                    <@\textcolor{red}{(*)}@>
\end{lstlisting}

\subsection{\textup {Lorenz extended system}}\label{b}
\begin{lstlisting}[belowskip=0pt]%[    basicstyle=\footnotesize, %or \small or \footnotesize etc.]

function f=Lorenz_ext(t,x)
f=zeros(size(x));
% ne*(ne+1) variables allocated for variational equations
% Here a=10, b=23, c=8/3;
X= [x(4), x(7), x(10);
    x(5), x(8), x(11);
    x(6), x(9), x(12)];%To be modified if ne>3
% ne equations (Lorenz system)
f(1)=10*(y-x);
f(2)=-x*z+23*x-y;
f(3)=x*y-8/3*z;
% Jacobian matrix
J=[-10,  10,  0;
   23-z, -1, -x;
     y,  x, -8/3];
% Variational equations
f(4:12)=J*X; % To be modified if ne>3

\end{lstlisting}

\end{multicols}
\newpage{\pagestyle{empty}}

\begin{figure}
\begin{center}
\includegraphics[scale=.85]{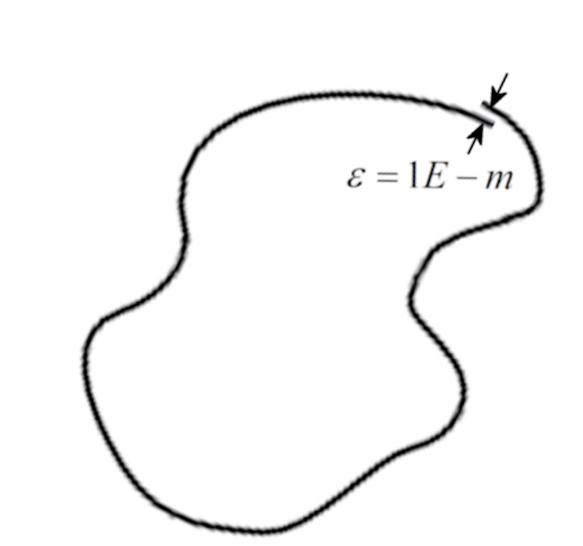}
\caption{Sketch of a NPT.  }
\label{fig2}
\end{center}
\end{figure}

\begin{figure}
\begin{center}
\includegraphics[scale=0.75]{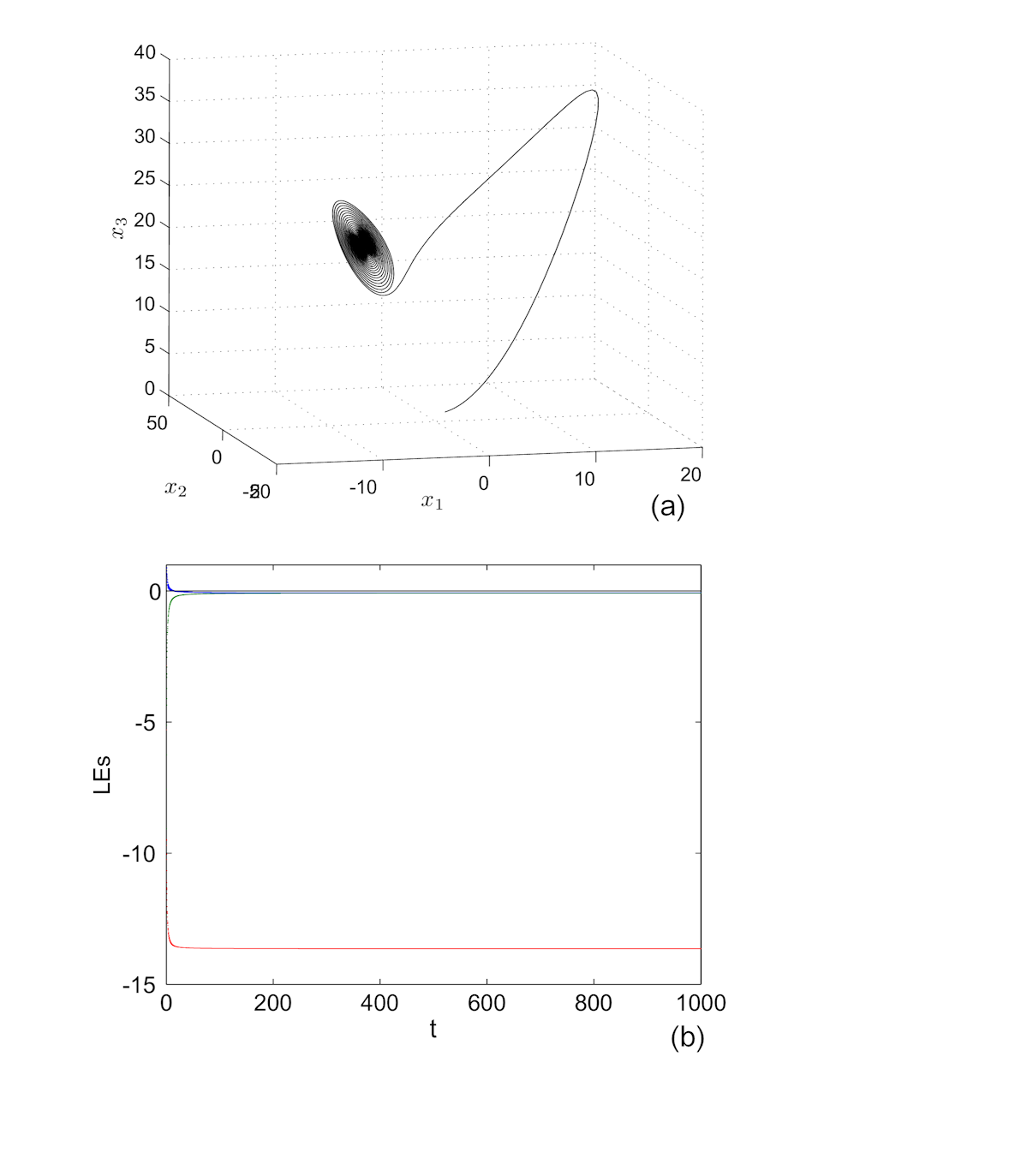}
\caption{(a) Phase plot of a trajectory of the Lorenz system \eqref{lorus} for $b=23$; (b) Time evolution of LEs for $b=23$.}
\label{fig3}
\end{center}
\end{figure}

\begin{figure}
\begin{center}
\includegraphics[scale=0.75]{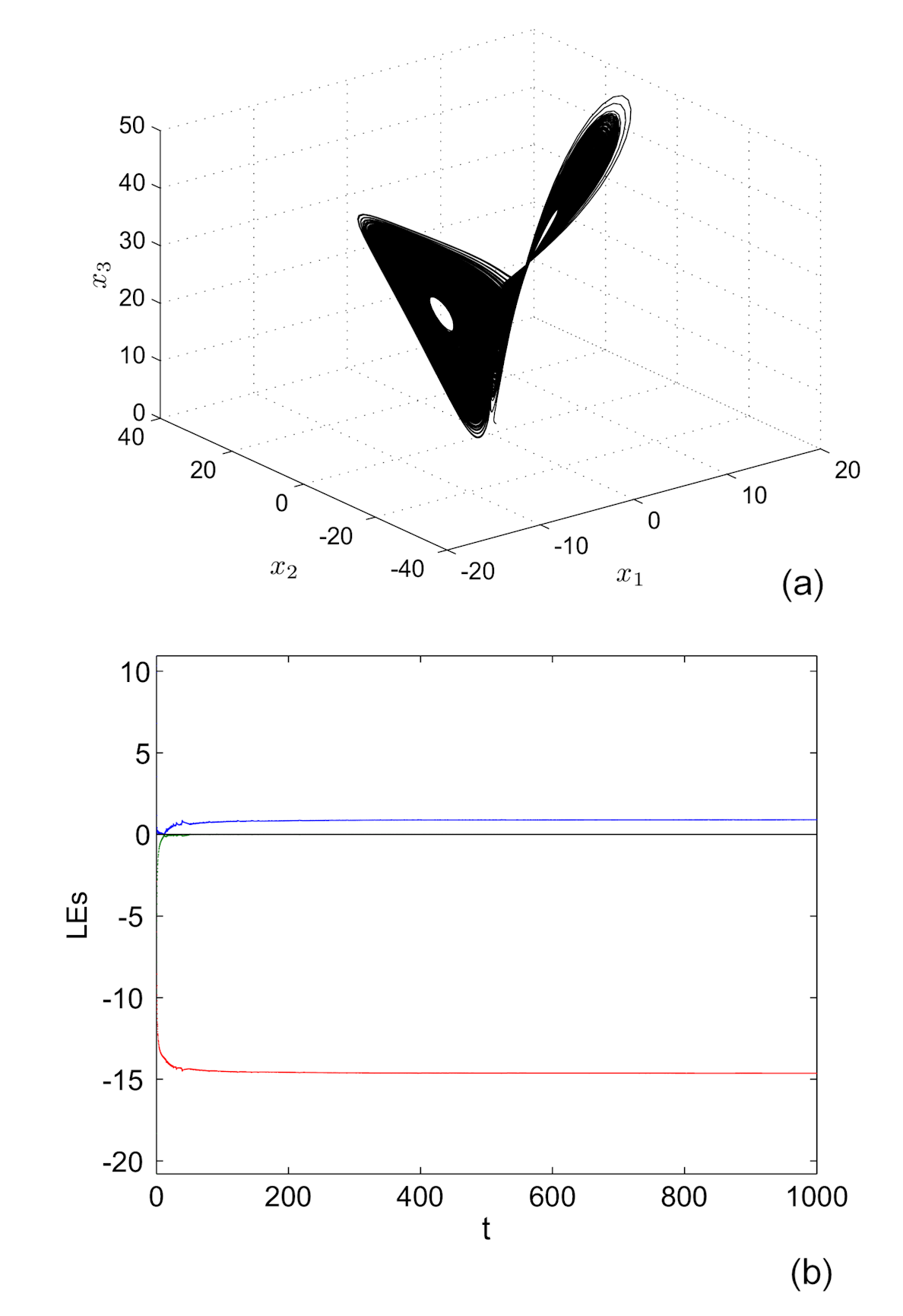}
\caption{(a) Phase plot of a trajectory of the Lorenz system \eqref{lorus} for $b=28$; (b) Time evolution of LEs for $b=28$.}
\label{fig4}
\end{center}
\end{figure}

\begin{figure}
\begin{center}
\includegraphics[scale=1]{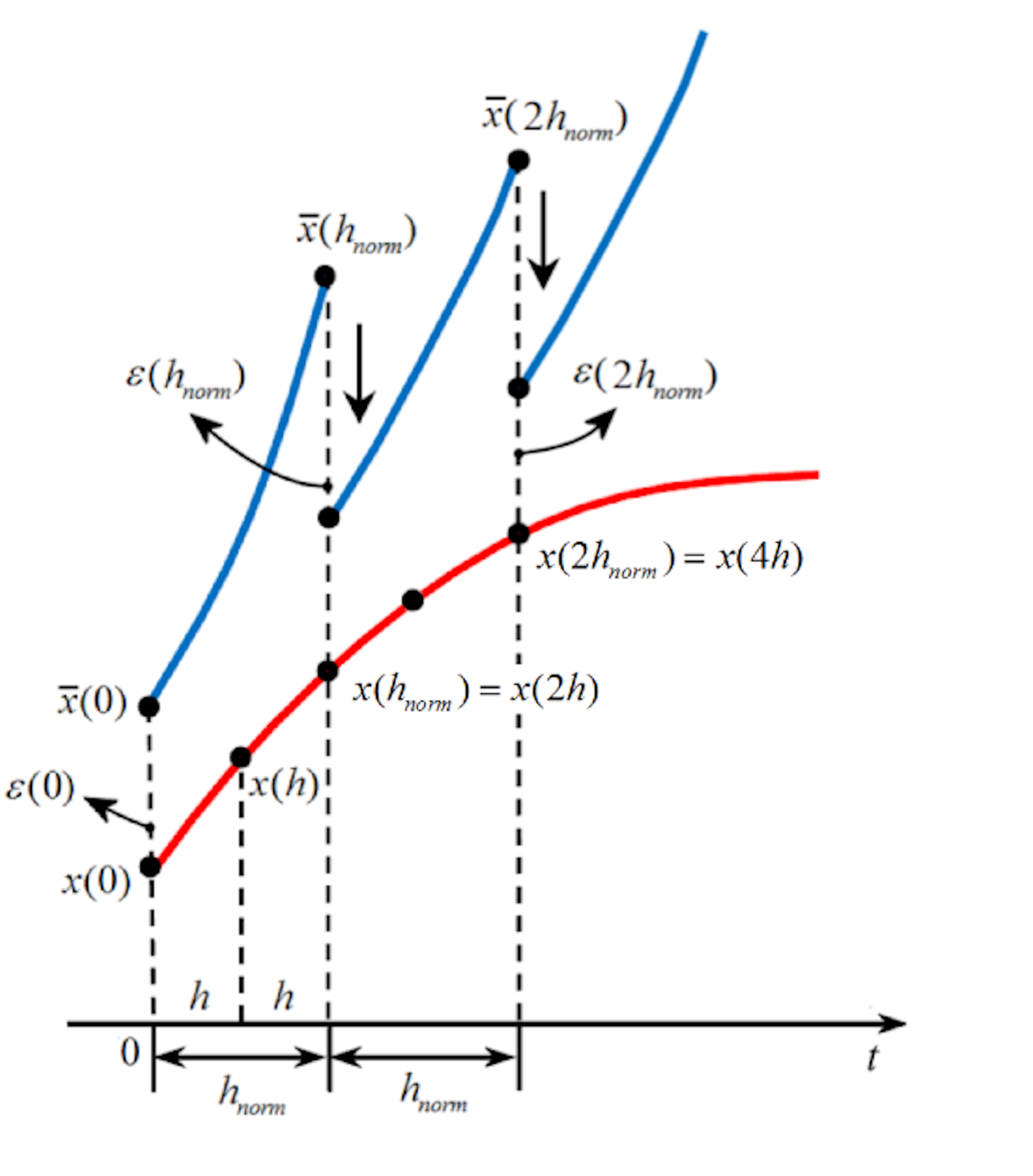}
\caption{Sketch showing the relation between $h_{norm}$ (Gram-Schmidt algorithm) and $h$ (step-size of the numerical integrator) \cite{dadus}.}
\label{fig5}
\end{center}
\end{figure}

\end{document}